\documentclass{article}
\usepackage{spconf,amsmath,graphicx,hyperref}
\usepackage{amssymb, booktabs}
\usepackage[justification=centering]{caption}
\usepackage{tikz}

\title{Audio-Visual Speech Enhancement for Spatial Audio - Spatial-VisualVoice and the MAVE Database}
\name{Danielle Yaffe$^{1}$, Ferdinand Campe$^{2}$, Prachi Sharma$^{2}$, Dorothea Kolossa$^{2}$, Boaz Rafaely$^{1}$\thanks{Funded by Deutsche Forschungsgemeinschaft (DFG, German Research Foundation) – Project number KO3434/9-1}}
\address{$^{1}$School of Electrical and Computer Engineering, Ben-Gurion University of the Negev \\ $^{2}$Technische Universität Berlin}
\begin{document}
%

\maketitle
\begin{abstract}
Audio-visual speech enhancement (AVSE) has been found to be particularly useful at low signal-to-noise (SNR) ratios due to the immunity of the visual features to acoustic noise. However, a significant gap exists in AVSE methods tailored to enhance spatial audio under low-SNR conditions. The latter is of growing interest with augmented reality applications. To address this gap, we present a multi-channel AVSE framework based on VisualVoice that leverages spatial cues from microphone arrays and visual information for enhancing the target speaker in noisy environments. We also introduce MAVe, a novel database containing multi-channel audio-visual signals in controlled, reproducible room conditions across a wide range of SNR levels. Experiments demonstrate that the proposed method consistently achieves significant gains in SI-SDR, STOI, and PESQ, particularly in low SNRs. Binaural signal analysis further confirms the preservation of spatial cues and intelligibility.
\end{abstract}
\section{Introduction}
\label{sec:intro}

Spatial audio refers to sound associated with three-dimensional auditory space, allowing listeners to localize and experience sound from specific directions and distances. Spatial audio is rapidly becoming an important part of immersive media, virtual reality (VR), and augmented reality (AR), through binaural reproduction of sound over headphones. High-quality multi-channel recordings are essential for reproducing realistic auditory scenes with spatial fidelity and perceptual immersion \cite{rafaely2022spatial, cobos2022overview}. However, recordings of real-world acoustic environments often contain multiple speakers, reverberation, and background noise, posing significant challenges for high-quality spatial reproduction. Furthermore, spatial audio is increasingly captured by embedded microphone arrays (e.g., wearable devices) \cite{madmoni2021binaural, lee2021multichannel} used for speech communication in a virtual or augmented audio setting, and so in addition to spatial audio reproduction, there is an increasing need for speech enhancement (SE), to remove interfering sounds from recorded speech before employing spatial reproduction.

The task of SE is well studied in the literature, typically using multiple channel recordings and deep neural networks \cite{rafaely2022spatial, cobos2022overview, huang2025advances, michelsanti2021overview}. SE tailored for spatial audio is an emerging field of interest, aiming to preserve spatial integrity and create realistically perceived auditory scenes — a capability that is crucial for VR and AR applications. For example, \cite{cornelis2010performance} suggests binaural beamforming focusing on preserving the spatial features of the target speaker. An extension was suggested by \cite{hadad2016extensions} to address preserving the spatial impression of the interfering source. However, these methods require the signals at the ears of the listeners, which restricts the choice of the microphone arrays that can be used. Spatial audio enhancement tailored for Ambisonics signals \cite{rafaely2015fundamentals} has also been proposed, but requires spherical microphone array configurations that are less applicable in edge devices \cite{lugasi2020speech}. Studies on arbitrary array configurations try to bridge this gap \cite{lugasi2022spatial, lior2020beamforming}. While current methods for spatial audio enhancement show good performance, they have not yet been evaluated for highly challenging conditions such as very low signal-to-noise ratios (SNRs). 

When the SNR is particularly poor, audio-only methods struggle. Incorporating visual information for SE tasks can improve the robustness and accuracy, since the visual information (e.g., lip movements, facial attributes) are not affected by the acoustic environment \cite{sumby1954visual, gu2020multi}. Together, the two modalities, audio and visual, complement each other, mimicking human perception and creating a foundation for superior SE. Transformer-based models, including dual-stream and graph-enhanced variants, have shown strong performance across audio-visual (AV) tasks, while convolutional and recurrent architectures are also widely used \cite{visualvoice_paper, ma2023auto, wahab2024multi, lin2023av}. However, most existing audio-visual speech enhancement (AVSE) models operate on single-channel audio, effectively collapsing spatial cues into a monaural representation. As a result, state-of-the-art AVSE systems cannot be directly applied to spatial audio, which relies on multi-channel recordings to ensure high-quality spatial realism.

This limitation has motivated growing interest in incorporating multi-channel approaches for AVSE that leverage inter-microphone relations. The work in \cite{gu2020multi, yu2020audio, shao2022multi} introduce inter-microphone phase differences (IPDs) computed by the phase difference between channels of complex spectrograms. While these studies clearly highlight the benefit of using multi-channel audio and lip video information, their main goal is automatic speech recognition and not human perception or spatial audio reproduction. Therefore, they are not designed to preserve spatial cues in the recorded signals. 

In summary, despite progress in AVSE, a significant gap remains in low SNR conditions for spatial audio. To the best of our knowledge, no prior work has explicitly tailored AVSE to preserve and enhance multi-channel spatial audio. In this paper, we propose an AV network designed for wearable microphone arrays that enhances the target speaker while preserving spatial cues across all channels in various SNR conditions. Our main contributions are threefold:
(1) We introduce \textbf{MAVe}, a novel multi-channel audio-visual database designed for AVSE tasks.
(2) We propose an AVSE framework tailored for spatial audio, under low-SNR conditions.  
(3) We demonstrate the effectiveness of the proposed approach in realistic multi-speaker scenarios simulated with wearable-like microphone arrays configuration, achieving significant improvements in SI-SDR, PESQ, and STOI, both for the microphone and the binaural signals.

\section{System Model}
\label{sec:system_model}
Let us consider an array of $M$ microphones that captures speech signals, $s_i(t),  i\in{\{1,..,I\}}$, where $I$ is the number of speakers. Without loss of generality, the target speaker is indexed as $i=1$. Let $h_{i,m}$ denote the room impulse response (RIR) from the $i$-th speaker to the the $m$-th microphone. The signal at microphone $m$ due to speaker $i$ can therefore be formulated as, $x_{i,m}(t) = s_i(t) \ast h_{i,m}(t)$, where $\ast$ indicates the convolution operation. Adding the contributions from all speakers, and including sensor noise $n_m(t)$, the signal at microphone $m$ can be formulated as:
\begin{equation}
\begin{aligned}
y_m(t) &= x_{1,m}(t) + \sum_{i=2}^{I} x_{i,m}(t) + n_m(t)
\end{aligned}
\end{equation}
where $m\in{\{1,2,..,M\}}$ and $t$ denotes time, $t \in \{1,..,T\}$. The signal captured by the microphones can be written in a vector form as: $\boldsymbol{y}(t) = [y_{1}(t),...,y_{M}(t)]^\mathrm{T}$ where $\mathrm{T}$ represents the transpose operator.

The signals at the microphones, without sensor noise, due to the desired speech source, can be written in a vector form as: $\boldsymbol{x}_1(t) = [x_{1,1}(t),...,x_{1,M}(t)]^\mathrm{T}$. Given the noisy signals captured by the microphone array, the task of AVSE tailored for spatial audio can be described as estimating the direct and reverberant component of the target speech signal at the microphones, $\boldsymbol{\hat{x}}_1(t) \in \mathbb{R}^{M}$, using visual information of the target's face and lip movements denoted by $\boldsymbol{v}_1(k) \in \mathbb{R}^{P_\mathrm{v}}$ with frame index $k\in\{1,...,K\}$ and pixel count $P_\mathrm{v}$. The enhancement model, $F$, maps AV inputs to an estimated signal, i.e., 
\begin{equation}
\hat{\boldsymbol{x}}_{1}(t) = F(t; \boldsymbol{y}(\cdot), \boldsymbol{v}_1(\cdot))
\end{equation}
Unlike conventional single-channel AVSE, this formulation explicitly preserves the target reverberant speech in all microphones, which is required for spatial audio reproduction as it isolates the target voice in each channel for every time point.

\section{MAVe Database}
\label{sec:mave_data}
Research in AV spatial audio enhancement for low SNR requires a dataset that contains multi-channel audio captured by microphone arrays, in particular wearable arrays, in realistic and various room scenarios with accompanying visual attributes for visual processing. One dataset existing is the EasyCom dataset \cite{donley2021easycom} however it does not offer controlled, reproducible conditions suitable for systematic AVSE studies. 

To fill this gap, we introduce MAVe (Multi-channel Audio-Visual, wearable) database. MAVe is a diverse multi-channel audio-visual dataset derived from the GRID corpus \cite{cooke2006audio}. The GRID corpus contains high and moderate quality video recordings of 33,000 semantically unpredictable sentences spoken by 33 speakers (18 male, 15 female). It was designed for AVSE and AV speech recognition training and testing, as well as for supporting research on acoustic and AV speech perception. Audio is sampled at 16 kHz and video at 25 fps. Each utterance from the clean single-speaker recordings is paired with synchronized video of the target speaker’s face.

To simulate realistic room scenarios, we introduce interfering speakers from the same corpus in addition to the target speaker, and spatialize all using Pyroomacoustics library \cite{scheibler2018pyroomacoustics} in random locations in a room setting. Multi-channel mixtures are generated with a four-microphone array mounted on a glasses-like wearable device, in free-field settings, simulating a recording in a realistic noisy room. RIRs are synthesized with varying room sizes: $7 \times 8 \times 3~\mathrm{m}^3$, $10 \times 8 \times 3~\mathrm{m}^3$, and $12 \times 9 \times 3~\mathrm{m}^3$, and reverberation times (T60) ranging from 0.2s to 1.0s. To increase realism, small temporal offsets are introduced between the target and interfering sources, reflecting the natural asynchrony of conversational speech that is absent in the GRID corpus. In addition, we ensure that the target speaker is present in a short segment without overlap, which is critical in a single-channel AVSE to identify the target speaker. The MAVe database covers a wide range of realistic low SNR conditions by mixing the target and interfering signals at controlled SNRs ranging from -10 dB to 5 dB. For evaluation, both the clean target (with room effects but no interference) and the noisy mixtures are preserved, allowing for the assessment of enhancement the quality and the preservation of spatial cues in the binaural rendering. 

The MAVe dataset will be made publicly available upon publication\footnote{A download link will be provided in the camera-ready version of this paper.}.

\section{Proposed Method}
\label{sec:proposed_method}

    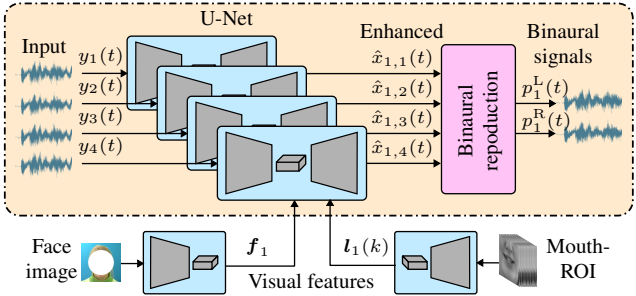
\begin{figure}
        \centering
        \begin{tikzpicture}
        \node[anchor=north west, inner sep=0] (img) {\includegraphics{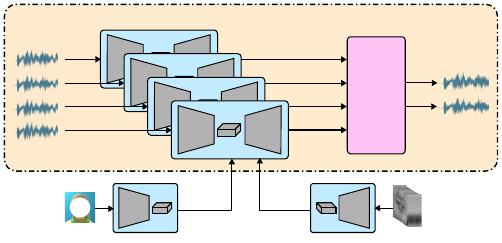}};
    
        \begin{scope}[x=1mm, y=-1mm]
          \node[anchor=north west, rounded corners, inner sep=2pt, font=\footnotesize]
            at (26,1) {U-Net};
          \node[anchor=north west, inner sep=0, font=\footnotesize]
            at (3,5.4) {Input};
          \node[anchor=north west, inner sep=0, font=\footnotesize]
            at (48,3.4) {Enhanced};
          \node[anchor=north west, text width=10mm, align=center, inner sep=0, font=\footnotesize]
            at (70,3.4) {Binaural\\signals};
          \node[anchor=north west, inner sep=0, font=\footnotesize]
            at (33.1,36.7) {Visual features};
        
          \node[anchor=north west, text width=17mm, align=center, inner sep=0, font=\footnotesize, rotate=90]
            at (61,25) {Binaural\\ repoduction};
          \node[anchor=north west,text width=10mm,    
  align=center, inner sep=0, font=\footnotesize]
            at (72,32) {Mouth-ROI};
          \node[anchor=north west,text width=10mm,    
  align=center, inner sep=0, font=\footnotesize]
            at (2,32) {Face image};
        
          \node[anchor=north west, inner sep=0, font=\scriptsize]
            at (45.8,32) {$\boldsymbol{l}_1(k)$};
          \node[anchor=north west, fill=white, inner sep=0, font=\scriptsize]
            at (32.8,32) {$\boldsymbol{f}_1$};

          \node[anchor=north west, inner sep=0, font=\scriptsize]
            at (10.5,6.5) {$y_1(t)$};
          \node[anchor=north west, inner sep=0, font=\scriptsize]
            at (10.5,10.5) {$y_2(t)$};
          \node[anchor=north west, inner sep=0, font=\scriptsize]
            at (10.5,14.5) {$y_3(t)$};
          \node[anchor=north west, inner sep=0, font=\scriptsize]
            at (10.5,18.5) {$y_4(t)$};
                    
          \node[anchor=north west, inner sep=0, font=\scriptsize]
            at (49.5,6.85) {$\hat{x}_{1,1}(t)$};
          \node[anchor=north west, inner sep=0, font=\scriptsize]
            at (49.5,10.85) {$\hat{x}_{1,2}(t)$};
          \node[anchor=north west, inner sep=0, font=\scriptsize]
            at (49.5,14.85) {$\hat{x}_{1,3}(t)$};
          \node[anchor=north west, inner sep=0, font=\scriptsize]
            at (49.5,18.85) {$\hat{x}_{1,4}(t)$};

          \node[anchor=north west, inner sep=0, font=\scriptsize]
            at (69.5,10.45) {$p^\mathrm{L}_{1}(t)$};
          \node[anchor=north west, inner sep=0, font=\scriptsize]
            at (69.5,14.85) {$p^\mathrm{R}_{1}(t)$};
        \end{scope}
      \end{tikzpicture}
    \caption{Spatial-VisualVoice model overview. Microphone signals are processed through a U-Net, with shared visual features. The enhanced signals are passes through the binaural reproduction block to produce binaural signals.}
    \label{fig:method}
\end{figure}

We extend the VisualVoice framework \cite{visualvoice_paper}, as depicted in Figure \ref{fig:method}, to a multi-channel setting through multiple single-channel networks, with adaptations that preserve spatial cues. A visual stream is used to extract lip features and facial attributes from a speaker video, which condition the processing of each microphone channel. Specifically, every audio channel is passed independently through a U-Net encoder–decoder, while sharing the same visual conditioning. This design allows the model to exploit the discriminative power of the target’s lip movements without collapsing the microphone signals into a single channel estimate, thereby maintaining natural inter-channel differences. The target speaker signal at each microphone is obtained by predicting a complex ideal ratio mask (cIRM) \cite{williamson2015complex}, $M_1$, and applying it to the mixture spectrogram:
\begin{equation}
\hat{x}_{1,m}(t) = \mathrm{ISTFT}\{Y_m(f,t) \cdot M_1(f,t)\},
\end{equation}
where $Y_m(f,t)$ is the short-time Fourier transform (STFT) of the signal at the microphone, $y_m(t)$ and ISTFT is the inverse STFT operation. For more architectural details, we refer the reader to \cite{visualvoice_paper}. The enhanced microphone signals are processed through a binaural rendering module, as seen in Figure \ref{fig:method} on the right, which reproduces the left-ear and right-ear signals, $p^\mathrm{L}(t)$ and $p^\mathrm{R}(t)$, respectively. This step produces natural binaural signals suitable for perceptual evaluation and downstream listening tasks.

\section{Experimental Study}
\label{sec:experimental_study}

\begin{table*}[t]
\centering
\caption{Speech enhancement results for microphone signals for SI-SDR, STOI, and PESQ across SNR levels and overall.}
\resizebox{\textwidth}{!}{%
\begin{tabular}{lccccccccccccccc}
\toprule
Model & \multicolumn{3}{c}{-10 dB} & \multicolumn{3}{c}{-5 dB} & \multicolumn{3}{c}{0 dB} & \multicolumn{3}{c}{5 dB} & \multicolumn{3}{c}{Overall} \\
 & SI-SDR & STOI & PESQ & SI-SDR & STOI & PESQ & SI-SDR & STOI & PESQ & SI-SDR & STOI & PESQ & SI-SDR & STOI & PESQ \\
\midrule
Noisy & -8.80 & 0.45 & 1.12 & -4.14 & 0.54 & 1.18 & 0.64 & 0.64 & 1.32 & 5.54 & 0.73 & 1.59 & -2.75 & 0.57 & 1.27 \\
VisualVoice & -5.98 & 0.54 & 1.22 & 0.10 & 0.65 & 1.44 & 5.52 & 0.74 & 1.81 & 8.92 & 0.80 & 2.25 & 0.95 & 0.66 & 1.60 \\
Audio-VisualVoice & 1.91 & 0.65 & 1.49 & 4.26 & 0.73 & 1.72 & 7.00 & \textbf{0.78} & 2.07 & \textbf{9.73} & \textbf{0.83} & 2.49 & 5.19 & 0.73 & 1.87 \\
Spatial-VisualVoice & 3.12 & \textbf{0.70} & \textbf{1.68} & 5.26 & \textbf{0.75} & \textbf{1.91} & 7.15 & \textbf{0.78} & \textbf{2.21} & 8.78 & 0.82 & \textbf{2.51} & 5.70 & \textbf{0.75} & \textbf{2.01} \\
SpatialQ-VisualVoice & \textbf{3.64} & \textbf{0.70} & 1.66 & \textbf{5.86} & \textbf{0.75} & \textbf{1.91} & \textbf{7.62} & \textbf{0.78} & 2.19 & 9.04 & 0.82 & 2.47 & \textbf{6.20} & \textbf{0.75} & 2.00 \\
\bottomrule
\end{tabular}}
\label{tab:persnr_results}
\end{table*}

\begin{table*}[t]
\centering
\caption{Speech enhancement results for binaural signals. SI-SDR, STOI, and PESQ averaged across both ear signals at different SNR levels and overall.}
\resizebox{\textwidth}{!}{%
\begin{tabular}{lccccccccccccccc}
\toprule
Model & \multicolumn{3}{c}{-10 dB} & \multicolumn{3}{c}{-5 dB} & \multicolumn{3}{c}{0 dB} & \multicolumn{3}{c}{5 dB} & \multicolumn{3}{c}{Overall} \\
 & SI-SDR & STOI & PESQ & SI-SDR & STOI & PESQ & SI-SDR & STOI & PESQ & SI-SDR & STOI & PESQ & SI-SDR & STOI & PESQ \\
\midrule
Noisy & -9.67 & 0.43 & 1.13 & -5.08 & 0.52 & 1.20 & -0.34 & 0.62 & 1.34 & 4.49 & 0.71 & 1.61 & -3.70 & 0.55 & 1.29 \\
VisualVoice & -7.00 & 0.53 & 1.24 & -1.04 & 0.63 & 1.45 & 4.45 & 0.72 & 1.82 & 7.89 & 0.79 & 2.23 & -0.07 & 0.65 & 1.61 \\
Audio-VisualVoice & 0.93 & 0.64 & 1.48 & 3.19 & 0.72 & 1.68 & 5.89 & \textbf{0.77} & 1.99 & \textbf{8.65} & \textbf{0.82} & 2.38 & 4.14 & 0.72 & 1.82 \\
Spatial-VisualVoice & 2.32 & \textbf{0.69} & \textbf{1.68} & 4.43 & \textbf{0.74} & \textbf{1.92} & 6.37 & \textbf{0.77} & \textbf{2.20} & 8.09 & 0.81 & \textbf{2.52} & 4.93 & \textbf{0.75} & \textbf{2.02} \\
SpatialQ-VisualVoice & \textbf{2.88} & \textbf{0.69} & 1.67 & \textbf{5.03} & \textbf{0.74} & 1.91 & \textbf{6.83} & \textbf{0.77} & 2.18 & 8.35 & 0.81 & 2.46 & \textbf{5.43} & 0.74 & 2.00 \\
\bottomrule
\end{tabular}}
\label{tab:persnr_results_bin}
\end{table*}

\subsection{Setup}
\label{sec:setup}
We use the MAVe database, with the moderate-quality visual stream providing a realistic yet memory-efficient approximation of recording conditions. For training the proposed method and studied variants are initialized from pre-trained VisualVoice weights and fine-tuned, leveraging rich audio–visual representations learned from large-scale pretraining. This not only accelerates convergence and improves stability but also provides an inherent tendency for associating visual cues with target speech, which is crucial for effective enhancement in the multi-channel low-SNR setting. The fine-tuning is done on a dedicated training subset from MAVe database.

We test and analyze our results on a subset of unseen data from the MAVe database dedicated for testing. We evaluate enhancement performance using signal-to-distortion ratio (SI-SDR) \cite{le2019sdr} as a quantitative measure, and Perceptual Evaluation of Speech Quality (PESQ) \cite{rix2001perceptual} and Short-Time Objective Intelligibility (STOI) \cite{taal2011algorithm} as perceptual metrics. Higher is better for all metrics. Together, these metrics allow for a broad evaluation of enhancement performance.

\subsection{Methodology}
\label{sec:methodology}
We build upon the VisualVoice setup, while conducting an examination of loss formulations to steer the model in the task of spatial enhancement. We present the baseline models and our proposed variants, highlighting their training setup and respective roles in the evaluation.

\textit{VisualVoice \cite{visualvoice_paper}} is a widely used audio-visual enhancement model that combines an audio encoder-decoder with a visual encoder extracting embeddings from lip motion. Audio and visual representations are aligned through cross-modal discriminative and reconstruction losses, allowing the model to associate speech features with the corresponding speaker’s visual cues. Our audio-only variant, \textit{Audio-VisualVoice}, follows the same architecture as VisualVoice but omits visual information during the fine-tuning stage by setting $\boldsymbol{l}_1(k)$ and $\boldsymbol{f}_1$ to zero, enabling assessment of the contribution of visual features. The proposed AVSE method,  \textit{Spatial-VisualVoice}, process each channel independently while leveraging the target speaker’s visual features. We refer the reader to Section~\ref{sec:proposed_method} for further details. Finally, we introduce \textit{SpatialQ-VisualVoice}, which extends the original VisualVoice reconstruction-driven approach further by incorporating an SI-SDR–motivated loss that directly measures waveform fidelity for each microphone channel:
\begin{equation}
\mathcal{L}_{\mathrm{SI\text{-}SDR}}(\hat{x}, x) =
\log_{10}\left(\frac{|\hat{x} - \alpha x|^2}{|\alpha x|^2} \right),
\quad
\alpha = \frac{\langle \hat{x}, x \rangle}{|x|^2}
\end{equation}
This loss enforces accurate time-domain reconstruction across channels, complementing the visual conditioning applied to each microphone in Spatial-VisualVoice.

\subsection{Results - Microphone Signals}
\label{ssec:resultsmic}




Table~\ref{tab:persnr_results} presents enhancement performance across various SNR levels and mean performance across the entire test dataset under the Overall column. The original VisualVoice network trained on non-reverberant speech provides only small enhancements across all SNR levels, demonstrating the importance of training on reverberant speech under realistic recording scenarios in order to preserve spatial cues and accurately reconstruct the multi-channel signal. The Audio-VisualVoice, which is fine-tuned to be able to cope with missing visual data, demonstrates surprisingly strong results. At 0 dB, it yield an SI-SDR gain of more than 6 dB relative to the noisy data. At -10 dB, reaches 1.91 dB, substantially outperforming the original VisualVoice baseline (-5.98 dB). This confirms that even a short isolated segment of the target speaker is sufficient for the model to generalize effectively and achieve strong enhancement without video. Spatial-VisualVoice surpassed Audio-VisualVoice, in various SNR conditions. Both Spatial-VisualVoice and its SI-SDR variant maintain consistently high performance across all conditions including very low SNRs, highlighting their ability to preserve spatial cues where it matters most. At -10 dB, Spatial-VisualVoice and SpatialQ-VisualVoice yield an SI-SDR gain of 11.9 dB and 12.4 dB, respectively, measured relative to the noisy baseline. Specifically the SpatialQ-VisualVoice variant outperform all other models in SI-SDR except 5 dB due to the loss explicitly enforces accurate reconstruction. Also PESQ and STOI increase substantially compared to the original VisualVoice, indicating a clear perceptual improvement.

\subsection{Results - Binaural Signals}
\label{ssec:results_bin}
To further assess performance, we analyze the binaural signals. Binaural reproduction is done using the binaural signal matching (BSM) method \cite{lior2020beamforming}, which has shown to perform high quality binaural reproduction for complex sound fields and arbitrary microphone arrays. This enables evaluation of both perceptual quality and intelligibility in listening setup. 

Table~\ref{tab:persnr_results_bin} reports SI-SDR, STOI, and PESQ for the binaural outputs across SNR levels and across the test dataset in the Overall column. Performance over SNRs is consistent with the microphone-signal analysis. SpatialQ-VisualVoice provide the largest gains in low-SNR conditions. For example, at -10 dB SNR, SpatialQ-VisualVoice attains 2.88 dB SI-SDR, compared to -7.00 dB for VisualVoice and 0.93 dB for Audio-VisualVoice. At 5 dB, all models improve as the task becomes less challenging. STOI and PESQ scores are consistently high for the proposed Spatial-VisualVoice indicating that integrating multi-channel spatial information with visual guidance consistently produces perceptually natural enhancements across SNR levels.

Informal listening tests further indicate that the binaural outputs using Spatial-VisualVoice maintain naturalness and spatial clarity, confirming the effectiveness of the proposed approach. More formal tests are planned for future work. 

\section{Conclusion}

We present Spatial-VisualVoice, a multi-channel AVSE framework designed to preserve spatial cues in low-SNR, reverberant conditions. Leveraging visual features of the target speaker while processing each microphone channel independently, our method achieves consistent improvements. We introduce the MAVe database, a multi-channel audio-visual dataset derived from GRID, enabling controllable and realistic training and evaluation for wearable microphone arrays. We conducted a detailed microphone and binaural signal evaluation, demonstrating that our method effectively maintains spatial accuracy and speech intelligibility. Experiments show that incorporating both spatial and visual information is crucial for high-quality, intelligible, and spatially coherent enhancement in low SNRs. Suggestions for future work include extension of Spatial-VisualVoice to a multiple-input network, and evaluation via listening and intelligibility tests. 
 

%
%


\clearpage
\vfill\pagebreak

\small
\bibliographystyle{IEEEbib}
\bibliography{strings,refs}

\end{document}